\documentclass[a4paper,12pt]{article}
\usepackage[T1]{fontenc}
\usepackage[utf8]{inputenc}
\pdfoutput=1
\usepackage{amsmath}
\usepackage{amsfonts}
\usepackage{amssymb}
\usepackage[english]{babel}
\usepackage[pdftex]{graphicx,color}
\begin{document}
\title{A non Abelian effective model for ensembles of magnetic defects in $QCD_{3}$}
\author{C.~D.~Fosco$^{a}$ and L.~E.~Oxman$^{b}$\\ 
{\normalsize\it $^a$Centro At\'omico Bariloche and Instituto Balseiro}\\ 
{\normalsize\it Comisi\'on Nacional de Energ\'\i a At\'omica}\\ 
{\normalsize\it R8402AGP Bariloche, Argentina.}\\
{\normalsize\it $^b$ Instituto de F\'{\i}sica}\\
{\normalsize\it Universidade Federal Fluminense}\\
{\normalsize\it Campus da Praia Vermelha}\\
{\normalsize\it Niter\'oi, 24210-340, RJ, Brazil.}}
\date{}
\maketitle
\begin{abstract}
We construct a non Abelian model for $SU(2)$ $QCD$ in Euclidean
three-dimensional spacetime and study its different phases. The model
contains a center vortex sector coupled to a dual effective field encoding
information about how the vortices are paired in the ensemble.  The
possible phases in parameter space are interpreted in terms of the
proliferation of either closed center vortices or closed chains, where the
endpoints of open vortices are attached in pairs to monopole-like defects. 
\end{abstract}
\section{Introduction}\label{sec:intro}
The vortex model introduced by t' Hooft~\cite{3} is a low energy effective
theory that successfully describes some aspects of the confinement
mechanism in $2+1$ dimensional $SU(N)$ Yang-Mills theories. It is defined
in terms of a dynamical variable which is a complex scalar field $V$
equipped with a discrete $Z(N)$ symmetry, realized with a Lagrangian,
\begin{equation}
	{\mathcal L} \;=\; \partial_\mu \bar{V} \partial_\mu V + \mu^2 \bar{V} V + \alpha (\bar{V} V)^2 + \beta (V^N +\bar{V}^N)\;.
\label{prim}
\end{equation}
Its form is based on a study of the possible nontrivial vortex correlation
functions in the original theory. In particular, the confining phase is
described as one where the discrete $Z(N)$ symmetry is spontaneously
broken, due to the presence of a vortex condensate.  The possibility of an
effective field representation for $3D$ center vortices relies on the fact
that  an ensemble of stringlike objects can be thought of as a sum over
different numbers of particle worldlines, which corresponds to a second
quantized field theory \cite{Ambjorn}-\cite{kleinert}.  Based on this
observation, one of us proposed, in refs.~\cite{lucho,ldual}, a generalized
vortex model where $\partial_\mu$ in eq. (\ref{prim}) is substituted by the
covariant derivative $D_\mu$, that depends on a dual vector field
$\lambda_\mu$ describing the off-diagonal sector.  The dynamics is
completed with a Proca action term for $\lambda_\mu$. In ref.~\cite{boa},
we derived this model by considering an ensemble of chains, where the
vortex endpoints are attached in pairs to monopole-like defects, and
following recent polymer techniques to compute the vortex end-to-end
probability.

Effective field models can also be obtained in scenarios based just on the
monopole (instanton) component.  In this case, the assumption of Abelian
dominance and the associated monopole ensemble is encoded in a sine-Gordon
type model for a scalar dual field (see \cite{antonov,koko} and references
therein), as occurs in the case of compact $QED(3)$, discussed by Polyakov in ref.  \cite{polya}.

In spite of the fact that the initial theory is a non Abelian one,
these effective models are {\em Abelian\/}, that is, additional information
regarding this transition is already incorporated, while it would be
desirable to see it appearing as a phase transition in a previous non
Abelian model. 

In addition, the different ideas regarding the magnetic sector have been
explored in the lattice, relying only on monopoles \cite{cp}-\cite{KLSW},
only on center vortices~\cite{debbio3}-\cite{quandt}, or on
chains~\cite{AGG}-\cite{GKPSZ}. Therefore, it would be interesting to
construct a model where the possible phases in parameter space correspond
to the different ensembles.

In this article, we construct a non Abelian effective model which
encompasses a description of interacting effective gluons and center
vortices. Depending on the choice of parameters, the vortices can be found
in different states, including a phase where they are closed, and a phase
where their endpoints become paired to form closed chains. 

To that aim, we use a parametrization that treats the different color
components in a symmetric way \cite{oxman2}, and describes correlated
monopoles and center vortices as defects of a local color frame
$\hat{n}_a$, $a=1,2,3$. This parametrization is based on the usual manner
to introduce thin center vortices in Yang-Mills theories
\cite{engelhardt1,reinhardt}, and corresponds to a symmetric form of the
Cho-Faddeev-Niemi (CFN) decomposition~\cite{cho-a}-\cite{Shaba}, used to
represent monopoles as defects of the third component $\hat{n}=\hat{n}_3$.
For a description of center vortices in the CFN framework, and related
consequences in the continuum, see refs. \cite{lucho,ldual}.

Since center vortices can be joined in pairs to pointlike monopoles, the
natural non Abelian field content of the model is given by a scalar field with 
one (magnetic) color index, generalizing the vortex field $V$ in eq. (\ref{prim}),
and a scalar field with two color indices, generalizing the scalar dual
field in scenarios only involving the monopole
component~\footnote{Interestingly, isospin two order parameters appear in
models for liquid crystals ~\cite{degennes-crys}.}.  The order parameters
present in the effective model bear a relation to the nature of the phase
transition one may describe. In this respect, the interesting point has
been raised~\cite{koko} about whether the confining/deconfining phase
transition is of the KT or Ising model type.  The former involves the
monopole sector: at high temperatures, the instanton magnetic flux is
distributed along the two spatial directions, thus leading to effective
logarithmic interactions. Then, because of dimensional reduction,
instantons and anti-instantons tend to be suppressed by forming pairs.  On
the other hand, the latter naturally involves the vortex degrees of
freedom, as they are the objects where the discrete symmetry
transformations act.

In the model we construct and study below, since it does contain order
parameters for both the center vortices and the distribution of
monopole-like defects they can concatenate to form chains, an interesting
framework to discuss the competition between different phases shall emerge,
originating a phase diagram with a rich structure.  
This paper is organized as follows: in section~\ref{sec:nonab}, we deal
with the topological defects included in the model, in particular, their
parametrization, and the functional and ensemble integration over them. In
section~\ref{sec:effective}, based on the previous section results, and after
discussing the possible symmetries, we construct an action for the
effective model in terms of the fields introduced therein.  Finally, in
section~\ref{sec:phase}, we present a study of the phase structure of the
model, based on some assumptions about the relative strength of its
different terms.
\section{Non Abelian defects in YM theories}\label{sec:nonab}
We shall start from the $SU(2)$ Yang-Mills action, $S_{YM}$, which may be written as follows:
\begin{equation}
S_{YM}\;=\;\frac{1}{4}\int d^3x\; \vec{F}_{\mu \nu}\cdot \vec{F}_{\mu
\nu}\;\;, 
\end{equation}
where ${\vec F}_{\mu\nu}$ is the non Abelian field-strength tensor. We use
an arrow on top of any object to denote the $3$-component vector formed by
its components on the $su(2)$ Lie algebra basis, whose elements are the
(Hermitian) generators $(T^a)_{a=1}^3$. In the concrete case we are
considering, they can be conveniently realized as $T^a=\tau^a/2$, where
$\tau^a$ denotes a Pauli matrix; they satisfy $\left[
T^a,T^b\right]\,=\,i\, \epsilon^{abc}\, T^c$, and ${\rm tr} (T^a
T^b)=\frac{1}{2}\delta^{ab}$.

Thus, with this notation, we may write down the defining equation for
${\vec F}_{\mu\nu}$, as follows: 
\begin{equation}\label{eq:deff}
\vec{F}_{\mu \nu} \cdot \vec{T}\, = \, \frac{i}{g} \, \left[D_\mu,D_\nu
\right] \;\;,\;\;\; D_\mu\,=\,\partial_\mu-ig \vec{A}_{\mu}\cdot \vec{T}
\;,
\end{equation}
where $D_\mu$ has been used to denote the covariant derivative operator,
when acting on fields in the fundamental representation.

It goes without saying that a `canonical color basis' $\big({\hat
e}^a\big)_{a=1}^3$ (with color components ${\hat e}^a_b = \delta_{ab}$) can
be introduced, so that $\vec{A}_\mu = \vec{A}_\mu^a \, {\hat e}^a$. This
seemingly trivial remark is made in order to highlight the next step;
namely, that one could have used a different basis.  Indeed, in order to
describe configurations with defects, in a symmetric way that
admits its extension to finite size objects, we
introduce a space-dependent color basis $(\hat{n}_a)_{a=1}^3$, related to
the original one by: $S T^a S^{-1}= \hat{n}_a \cdot \vec{T}$ (with \mbox{$S
\in SU(2)$}). Thus the new basis is connected to the  canonical one by an
orthogonal space-dependent matrix $R(S)$: $\hat{n}_a=R(S)\hat{e}_a$, which
belongs to the adjoint representation. In this representation, the
corresponding infinitesimal generators shall be denoted by $M^a$, with
$(M^a)^{bc}\equiv -i \epsilon^{abc}$. They satisfy $\left[
M^a,M^b\right]=i\epsilon^{abc}M^c$, ${\rm tr}\,(M^a M^b)= 2 \delta^{ab}$. 
At this point, and equipped with the local basis, we consider the
parametrization of the gauge field \cite{oxman2}:
\begin{equation}
\vec{A}_\mu =({\cal A}^a_\mu-C^a_\mu)\, \hat{n}_a  ,
\label{ansatz}
\end{equation}
where the frame dependent fields,
\begin{equation}
C^a_\mu = -\frac{1}{2 g} \epsilon^{abc} \hat{n}_b \cdot \partial_\mu
\hat{n}_c \;,
\label{CAm}
\end{equation}
satisfy the properties: 
\begin{equation}\label{equali}
\hat{n}_b \cdot \partial_\mu \hat{n}_c = -g \epsilon^{abc} C^a_\mu
\;,\;\;
C^a_\mu \, M^a = \frac{i}{g}R^{-1}\partial_\mu R  \;.
\end{equation}
This corresponds to a symmetric form of the Cho-Faddeev-Niemi (CFN)
decomposition \cite{cho-a}-\cite{Shaba}.  In terms of the parametrization
(\ref{ansatz}) of the gauge field, we note that the field-strength tensor
becomes:
\begin{equation}
\vec{F}_{\mu \nu}\,=\,G^a_{\mu \nu}\, \hat{n}_a
\;,\;\;\; G^a_{\mu \nu} = {\cal F}^a_{\mu \nu}({\cal A})-{\cal F}^a_{\mu \nu}(C),
\label{geA}
\end{equation}
with \mbox{${\cal F}^{a}_{\mu \nu }({\cal A})\equiv\partial_{\mu }{\cal
A}_{\nu}^a-\partial _{\nu }{\cal A}_{\mu}^a + g \epsilon^{abc}{\cal A}_{\mu
}^b{\cal A}_{\nu }^{c}$} (and an analogous expression for ${\cal
F}^{a}_{\mu \nu }(C)$), while the Yang-Mills action is given by:
\begin{equation}\label{eq:defsym}
S_{YM}=\int d^3x\, \frac{1}{4} G^a_{\mu \nu}G^a_{\mu \nu} \;.
\end{equation}
Regarding the color components of the frame-dependent tensor ${\cal
F}^a_{\mu\nu}(C)$, they can also be obtained by commuting covariant
derivatives in the adjoint representation:
\begin{equation}
{\cal F}^a_{\mu \nu}(C)\, M^a\,=\, \frac{i}{g} \big[{\cal D}_\mu, {\cal
D}_\nu \big]\;, \;\; {\cal D}_\mu \equiv \partial_\mu-ig C^a_\mu M^a \;\;, 
\end{equation}
so that the second equality in (\ref{equali}) implies the alternative
expression for ${\mathcal F}^a_{\mu\nu}(C)$:
\begin{equation}
{\cal F}^a_{\mu \nu}(C) = \frac{i}{2 g} \, {\rm tr}\, (M^a
R^{-1}[\partial_\mu,\partial_\nu]R) \;.
\label{FC}
\end{equation}
This equation highlights the meaning of ${\cal F}^a_{\mu \nu}(C)$, by
showing that it can only be different from zero where $R$ has defects;
these, are characterized here by the noncommutativity of the mixed partial
derivatives. These defects are zero measure objects; in other words, the
partial derivatives will fail to commute on zero measure regions.  Being
this an effective theory, this should be interpreted as the assumption that
the model describes physics at distances much larger than the size of the
defects. 

Of course, there are infinitely many different local frames, and
corresponding fields ${\cal A}^a_\mu$, that can be used to describe one and
the same gauge field configuration, $A^a_\mu$. One can use that large
amount of freedom in order to split it into its `regular' and `singular'
parts.  Indeed, the ${\cal A}^a_\mu$ measure will represent topologically
trivial fluctuations. The singular configurations,
described by the frames, will have a measure representing an ensemble
integration over defects. 

In ref.~\cite{oxman2}, one of us has shown that the configuration in
(\ref{ansatz}) is tantamount to the usual way \cite{engelhardt1,reinhardt} to introduce thin center
vortices on top of a trivial field configuration ${\cal A}^a_\mu\,
\hat{e}_a$, namely, 
\begin{equation}
\vec{A}_\mu \cdot \vec{T} = S \vec{{\cal A}}_\mu \cdot \vec{T} S^{-1}+
\frac{i}{g} S\partial_\mu S^{-1} -\vec{I}_\mu(S)\cdot \vec{T}\;.
\label{thin}
\end{equation}
Because of the presence of the last term, this is not just a gauge
transformation of the topologically trivial gauge field.  Indeed, the
$\vec{I}_\mu(S)$ field corresponds to the so called ideal center vortex,
and is localized on a hypersurface $\Sigma$. This is the region which, when
traversed, makes $S$ change by a center element.  It is designed to cancel
the contribution in the second term  originated from the discontinuity of
$S^{-1}$, only leaving the effect of the border of $\Sigma$ where the thin
center vortices are located. That is, we can write, \begin{equation}
\frac{i}{g} S\partial_\mu S^{-1}|_{\Sigma} =\vec{I}_\mu(S)\cdot \vec{T},
\end{equation}
where the subscript in the left-hand side amounts to just keeping in the
calculation the term originated from the derivative of the discontinuity 
in $S^{-1}$.  Considering two regular mappings $U$, $\tilde{U}$, the ideal
vortex satisfies,
\begin{equation}
\vec{I}_\mu(US\tilde{U}^{-1})\cdot \vec{T}= U \vec{I}_\mu(S)\cdot \vec{T} U^{-1},
\label{iv-t}
\end{equation}
obtained from $\partial_\mu (\tilde{U}^{-1} S^{-1} U^{-1})|_{\Sigma}=
\tilde{U}^{-1}\partial_\mu S^{-1}|_{\Sigma}\, U^{-1}$, as the term
localized on $\Sigma$ is only generated when $\partial_\mu$ acts on
$S^{-1}$. The gauge field $\vec{A}_\mu=\vec{A}_\mu(\vec{\cal A},S)$ in (\ref{thin})
enjoys the following properties, 
\begin{equation}
\vec{A}^U(\vec{\cal A},S) = \vec{A}(\vec{\cal A},US) 
\makebox[.5in]{,}
\vec{A}(\vec{\cal A}, S) = \vec{A}(\vec{\cal A}^{\tilde U},
S\tilde{U}^{-1})\;.
\label{reg-t}
\end{equation}
Then, in terms of the $\vec{\cal A}$, $S$ variables we have a double
redundancy, the usual one associated with invariance of the Yang-Mills
action under gauge transformations, $\vec{A}^U_\mu \cdot \vec{T} = U
\vec{A}_\mu \cdot \vec{T} U^{-1}+ \frac{i}{g} U\partial_\mu U^{-1}$,
represented by $S\rightarrow US$, and other originated from the different
ways to express the same vector field, combining the transformation
$\vec{\cal A}^{\tilde U}_\mu \cdot \vec{T} = \tilde{U} \vec{{\cal A}}_\mu
\cdot \vec{T} \tilde{U}^{-1}+ \frac{i}{g} \tilde{U}\partial_\mu
\tilde{U}^{-1}$, together with a right multiplication of $S$.
 
At this point, we would like to emphasize that a nonperturbative definition
of the path integral in Yang-Mills theory is still lacking.  This comes
about as a gauge fixing procedure generally leads to Gribov copies \cite{gribov} in
that regime, so that it is difficult to define an appropriate object where
each physical situation is counted only once. The restriction to the
modular region has been usually implemented by means of the Zwanzinger
action \cite{zwanziger}. In this framework, in the infrared regime, the
path integral has been shown to be dominated by configurations near the
Gribov horizon. On the other hand, as is well-known, configurations
containing magnetic objects proliferate at the
horizon \cite{Bruck}-\cite{maas}.  From this perspective, it is natural to
fix the redundancy by introducing the identity $1=\Delta_{FP}[{\cal A}]
\int [d\tilde{U}]\, \delta[f({\cal A}^{\tilde U})]$, in the perturbative
sector where the Faddeev-Popov procedure is well defined.  In addition, as
the $S$ sector parametrizes correlated monopoles and center vortices, it
represents configurations at the horizon, relevant to describe the large
distance physics. Giving a configuration $S$, gauge fixing amounts to
choose a representative of the orbit $US$. Any condition imposed on
$\vec{A}(\vec{\cal A}, S)$ will be invariant under the
$\tilde{U}$-transformations in eq. (\ref{reg-t}).  This is also the case
for conditions depending on $\vec{I}(S)$, as it is invariant under right
multiplication (cf. eq. (\ref{iv-t})).

In this article we shall not attempt to derive a precise construction of
the integration measure. Rather, having the previous remarks, notation,
and conventions in mind, we argue that it is quite natural to propose the
following path integral, 
\begin{equation}
Z_{YM}= \int [d{\cal A}] [dS]\, \Delta_{FP}[{\cal A}] \delta[f({\cal A})]
e^{-S_{YM}[A]} \;,
\label{ZYM}
\end{equation}
where $dS$, represents the ensemble integration over monopoles and thin
center vortices, that is supposed to include its own appropriate gauge
fixing condition. Note that, in the trivial sector, where $S=S_r$ is
regular, we have $\vec{A}(\vec{\cal A},S_r)=\vec{\cal A}^{S_r}$ and the
associated contribution to (\ref{ZYM}) is the usual, perturbative one.
Only in that sector $\vec{A}(\vec{\cal A}^{\tilde U}, S)=\vec{A}(\vec{\cal
A}, S\tilde{U})$ may be identified with a gauge transformation.

As a final step, and as a guide to the construction of the effective model, we
rewrite the partition function in the equivalent form:
\begin{equation}
Z_{YM}=\int [d{\cal A}] [dS][d\lambda]\, \Delta_{FP}[{\cal A}]
\delta[f({\cal A})] e^{-\int d^3x\, \left[\frac{1}{2} \lambda^a_\mu
\lambda^a_{\mu}+i \lambda^a_{\mu} ({\cal F}^a_{\mu}({\cal A})-{\cal
F}^a_{\mu}(C))\right]},
\label{YM-1}
\end{equation}
where ${\cal F}^a_{\mu}=\frac{1}{2}\epsilon_{\mu\nu\rho}{\cal
F}^a_{\nu\rho}$, and we have introduced a color-valued auxiliary 
field $\lambda^a_\mu$ to deal with a first-order version of (\ref{eq:defsym}).

\section{The effective theory}\label{sec:effective}
Let us now derive a non Abelian effective field theory for the sector of
defects. The derivation will become possible by relying
on the symmetries exhibited by the ensemble integration. This effective theory shall contain mass parameters, which we assume are originated from those present in a 
(phenomenological) ansatz for
the action of the defects. In this regard, we note that up to now we have considered thin center vortices, parametrized as in (\ref{ansatz}).
However, lattice simulations~\cite{debbio3} point to the idea that they become thick objects, characterized by some finite radius of the order of 1fm.
Moreover, as discussed in~\cite{oxman2}, the stable objects in the continuum could 
in fact correspond to some deformation of the thin objects given in~(\ref{ansatz}), where the ``thin'' quantities $C_\mu^a$ are replaced by some smooth 
finite radius profiles ${\cal C}_\mu^a$. If this is assumed to be the case, rather than eqs. (\ref{geA}), (\ref{eq:defsym}), the Yang-Mills action would have the form,
\begin{equation}
S_{YM}=\int d^3x\, \frac{1}{4} ({\cal F}^a_{\mu \nu}({\cal A})-{\cal F}^a_{\mu \nu}({\cal C}))^2 + {\cal R}\;,
\label{YMthick}
\end{equation}
where ${\cal R}$ vanishes for thin center vortices. Note that the first term can be linearized, as we did before, by introducing the fields $\lambda_\mu^a$.
Besides, at large distances, approximating ${\cal C}_\mu^a$ by $C_\mu^a$, this term shall originate the terms appearing in the exponent of eq. (\ref{YM-1}), when the center vortices were considered to be thin. On the other hand, the second term (${\mathcal R}$), 
will be concentrated on the center vortices and at large distances will produce instead
an additional action $S_d$ for the defects. Therefore, in the general case, the ensemble integration must be written in the form, 
\begin{equation}
e^{-S_{v,m}[\lambda]}=\int [dS]\, e^{-S_d + i\int d^3x\, \lambda^a_{\mu} {\cal
F}^a_{\mu}(C)}\;.
\label{Svm}
\end{equation}
The second term in the exponent above has a local $SO(3)$ symmetry under right multiplication: $S \rightarrow S \tilde{U}$, changing the color basis
from $\hat{n}_a \cdot \vec{T} =S T^a S^{-1}$ to $\hat{n}'_a \cdot \vec{T}
=S\tilde{U} T^a \tilde{U}^{-1} S^{-1}$, that is,
$\hat{n}'_a=R(S)R(\tilde{U})\hat{e}_a$. 
Note that, using (\ref{FC}), and that $R(\tilde{U})$ contains no defects, we have,
\begin{equation}
{\cal F}^a_{\mu \nu}(C') = \frac{i}{2 g} \, {\rm tr}\, (R(\tilde{U})M^aR^{-1}(\tilde{U}) R^{-1}(S)[\partial_\mu,\partial_\nu]R(S)).
\label{utilde}
\end{equation}
In other words, a regular local rotation of $\lambda^a_\mu$ can be
translated to a regular local transformation of $S$. Then, if $S_d$ were
nullified, that is, if we were dealing with strictly thin center vortices,
$S_{v,m}[\lambda]$ would be invariant under local $SO(3)$ rotations, as the
transformation $S \rightarrow S \tilde{U}$ could be  absorbed by the
integration measure $dS$. In this regard, we would like to underline that
this measure is to be accompanied by an appropriate gauge fixing condition
that is invariant under right multiplication (see the discussion at the end
of the previous section).  However, in $S_{v,m}[\lambda]$, that symmetry
will be broken to a global one because of the thick character expected for
center vortices. To have a simple picture about this statement we note that 
an action for thick center vortices will typically contain a Nambu-Goto term plus other terms describing the center vortex rigidity 
\cite{engelhardt1,BPZ}. These pieces can be generated, for instance, from a large distance approximation
of the more symmetric term (in color space),  
\[
\int d^3x\, d^3y\, {\cal F}^a_{\mu \nu}(C)|_x\, G_M (x-y)\, {\cal F}^a_{\mu \nu}(C)|_y ,
\]
where $G_M$ is a kernel localized on a scale $1/M$. Now, as we have seen in eq. (\ref{utilde}), the field strength 
${\cal F}^a_{\mu \nu}(C)$ will rotate under local $\tilde{U}$ transformations. Therefore, as for any finite $M$ 
the integrand above depends on the field strength at different spacetime points, it will change under the local 
transformations, only leaving a symmetry under the global ones.

Based on purely geometrical/mathematical grounds,  the
possible kinds of defects can be straightforwardly classified as follows:
\begin{enumerate}  
\item[i)]Closed center vortices. 
\item[ii)]Monopoles and antimonopoles, joined by center vortices (each
pointlike object is joined by a pair of center vortices).  
\item[iii)]A particular limit of ii): A coincident pair of center vortices,
which should correspond to an unobservable Dirac string. 
\end{enumerate}
To proceed, let us consider a type ii) configuration (correlated monopoles
and center vortices). To that end we recall that, in refs.~\cite{lucho}-\cite{boa}, we have
considered a particular case of that situation, namely, when the
center vortex color points along the (locally) diagonal direction
$\hat{n}_3$. In that case, the effective field describing these objects
corresponds to a complex vortex field $V$.  In particular, in ref. \cite{boa}, we have shown how the ensemble integration 
over open center vortices, whose endpoints are joined in pairs to form closed chains, leads to an Abelian $Z(2)$ effective theory that 
can be written in terms of $V$, thus making contact between the initial 
representation and the final effective field theory. For this aim, we applied recent polymer techniques \cite{fred,fredbook} to deal with the end-to-end probability 
associated with center vortices interacting with a general vector field $\lambda_\mu$, and a scalar field needed to represent vortex-vortex interactions. 
However, it is far from straightforward to extend this type of derivation to the non Abelian context. Therefore, in our case, we will propose a model 
relying on the symmetries displayed by the initial representation, that strongly constraints the possible associated effective theories.

In our case, the candidate for a
vortex field has to be a real \mbox{$3$-component} field $\phi^a$
($a=1,2,3$), because of the global $SO(3)$ symmetry of the action
$S[\lambda]$. We shall also introduce an
isospin-$2$ field $Q$, where $Q$ is a traceless symmetric $3\times
3$ real matrix, encoding information about how the monopole-like defects that 
center vortices can concatenate are distributed. We may then consider in the effective theory, an invariant
term $V_I$ that couples the monopole and vortex sectors:
\begin{equation}
V_I \,=\, \zeta \,\phi^T Q\, \phi \;, \;\;\;\zeta \equiv  {\rm constant}\;
\label{int}
\end{equation}
which is invariant under the local $SO(3)$
transformations: 
\begin{equation}\label{eq:local}
\phi(x) \rightarrow R(x)\,\phi(x) \;,\;\; Q(x)\rightarrow R(x)\, Q(x)\,
R^T(x)\;.
\end{equation}

There are also invariant terms involving just either the vortex or the
monopole field. Regarding the former, we may include a `potential' term
$V_\phi$, with the general structure:
\begin{equation}
V_\phi \,=\, \frac{\mu^2}{2} \phi^T \phi +\frac{\lambda}{4}(\phi^T \phi)^2,
\end{equation}
where $\mu$ and $\lambda$ are arbitrary constants. On the other hand, for
the case of the monopole field, we recall that an order parameter $Q$, with
a similar structure,  is well-known in the context of liquid crystals.
Thus, we expect the relevant terms in the effective theory to be of the
same kind, namely, we may include a potential $V_Q$ \cite{degennes-crys}:
\begin{equation}
V_Q \,=\, \frac{A}{2} \delta + \frac{B}{3} \Delta + \frac{C}{4} \delta^2 +
\frac{D}{5} \delta \Delta + \frac{E}{6} \Delta^2+F \delta^3, \label{VQ}
\end{equation}
where $A,\ldots,F$ are constants, and we have introduced two independent
$SO(3)$ invariants~\footnote{Being a traceless real symmetric matrix, the
invariant content of $Q$ can be generated by two real invariants. For
example, two of its eigenvalues.} that can be built in terms of $Q$:
\begin{equation}
\delta = Tr\, Q^2 \;\;,\;\;\; \Delta = Tr\, Q^3\;.
\end{equation}

Thus, the three terms $V_\phi$, $V_Q$ and $V_I$ have the local symmetry
(\ref{eq:local}). This local symmetry will be broken to its global
counterpart by the kinetic terms; however, these terms shall be constructed
in such a way that they are compatible with a local discrete gauge
symmetry. This symmetry must be present, at least in a phase where the
vacuum is symmetric (no spontaneous symmetry breaking). 

In this regard, the field strength tensor ${\cal F}^a_{\mu}(C)$ can be
written as,
\begin{eqnarray}
{\cal F}^a_{\mu}(C) &=& \frac{1}{2} \epsilon_{\mu \nu \rho}\, {\cal F}^a_{\nu \rho}(C)\nonumber \\
 &=&  \epsilon_{\mu \nu \rho}\, \partial_\nu C^a_\rho + \frac{g}{2}
\epsilon_{\mu \nu \rho}\, \epsilon^{abd} C^b_\nu C^d_\rho ,
\end{eqnarray}
where,
\begin{eqnarray}
C^1_\mu &=& -\frac{1}{g} \hat{n}_2 \cdot \partial_\nu \hat{n}_3 \nonumber \\
C^2_\mu &=& -\frac{1}{g} \hat{n}_3 \cdot \partial_\nu \hat{n}_1 \nonumber \\
C^3_\mu &=& -\frac{1}{g} \hat{n}_1 \cdot \partial_\nu \hat{n}_2 \;\;. 
\end{eqnarray}

We will show that ${\cal F}^a_{\mu}(C)$ can be rewritten as,
\begin{equation}
{\cal F}^a_{\mu}(C) = \tilde{h}^a_\mu - h^a_\mu,
\label{mono}
\end{equation}
\begin{equation}
\tilde{h}^a_\mu=\epsilon_{\mu \nu \rho}\, \partial_\nu C^A_\rho 
\makebox[.5in]{,}
h^a_\mu= - \frac{1}{2g} \epsilon_{\mu \nu \rho}\, \hat{n}_a \cdot
(\partial_\nu \hat{n}_a\times \partial_\rho \hat{n}_a ) ,
\label{mono1}
\end{equation}
where, in the second tensor, no summation over $a$ is understood. 

Let us take, for example, the third component of the field strength tensor,
\begin{equation}
{\cal F}^3_{\mu} = \tilde{h}^3_\mu - h^3_\mu,
\end{equation}
\begin{equation}
\tilde{h}^3_\mu=\epsilon_{\mu \nu \rho}\, \partial_\nu C^3_\rho 
\makebox[.5in]{,}
h^3_\mu= - g \epsilon_{\mu \nu \rho}\, C^1_\nu C^2_\rho .
\label{ht3}
\end{equation}
In order to show that
\begin{equation}
h^3_\mu = - \frac{1}{2g} \epsilon_{\mu \nu \rho}\, \hat{n}^3 \cdot (\partial_\nu \hat{n}^3 \times \partial_\rho \hat{n}^3 ),
\label{h3}
\end{equation}
we can simply note that,
\begin{eqnarray}
\partial_\nu \hat{n}_3 & = & (\hat{n}_1 \cdot \partial_\nu \hat{n}_3) \hat{n}_1 + (\hat{n}_2 \cdot \partial_\nu \hat{n}_3) \hat{n}_2
+  (\hat{n}_3 \cdot \partial_\nu \hat{n}_3) \hat{n}_3 \nonumber \\
 & = & (\hat{n}_1 \cdot \partial_\nu \hat{n}_3) \hat{n}_1 + (\hat{n}_2 \cdot \partial_\nu \hat{n}_3) \hat{n}_2 \nonumber \\ 
 & = & g ( C^2_\nu\, \hat{n}_1 -  C^1_\nu\, \hat{n}_2 ).
\end{eqnarray}
Then, replacing in the second member of (\ref{h3}), and using
$\hat{n}_1 \times \hat{n}_2 = \hat{n}_3$, etc., it is straightforward
to make contact with (\ref{ht3}).

The important point is that (\ref{mono}) and  (\ref{mono1}) imply that
for a fixed monopole background correlated with center vortices, the
integral of each component over a closed surface $\partial \vartheta$
(given as the border of a three-volume $\vartheta$),
\begin{equation}
\oint_{\partial\theta} dS_\mu\, {\cal F}^a_{\mu}(C) =
\oint_{\partial\theta} dS_\mu\, (\tilde{h}^a_\mu - h^a_\mu)= \frac{1}{2g}
\oint_{\partial\theta} dS_\mu\,  \epsilon_{\mu \nu \rho}\, \hat{n}_a
\cdot (\partial_\nu \hat{n}_a \times \partial_\rho \hat{n}_a),
\end{equation}
gives the $\Pi_2$ topological charge for the mapping $\partial \vartheta \to \hat{n}_a$. More precisely,
\begin{equation}\label{eq:quantiz}
\oint_{\partial\theta} dS_\mu\, {\cal F}^a_{\mu}(C) = \frac{4\pi}{g} (n_+(\vartheta)-n_-(\vartheta)) ,
\end{equation}
where $n_+(\vartheta)$ ($n_-(\vartheta)$) is the number of monopole
(antimonopole) defects inside $ \vartheta $, for the component $\hat{n}_ a$.

It may appear that the previous expression sets a preferred direction in
color space. This impression can be dispelled by considering the effect
that a space independent change of color basis has on the expression
(\ref{eq:quantiz}). To that end, we consider a new basis $\big({\hat
n}'_a\big)_{a=1}^3$, related to the original one by a (constant) matrix
$R'$, namely: $\hat{n}'_a= R'\hat{n}_a$, $a=1,2,3$. In components, and
using an obvious notation, the last relation means:
\begin{equation}
(\hat{n}'_a)_b\;=\; (R')_{bc} (\hat{n}_a)_c \;.
\end{equation}
Thus, we see that:
$$
\oint_{\partial\theta} dS_\mu\,  \epsilon_{\mu \nu \rho}\, \hat{n}'_a
\cdot (\partial_\nu \hat{n}'_a \times \partial_\rho \hat{n}'_a)\,=\,
\oint_{\partial\theta} dS_\mu\,  \epsilon_{\mu \nu \rho}\,
\epsilon_{b_1b_2b_3} (\hat{n}'_a)_{b_1} 
\partial_\nu (\hat{n}'_a)_{b_2} \partial_\rho (\hat{n}'_a)_{b_3}
$$
$$
\,=\, \oint_{\partial\theta} dS_\mu\,  \epsilon_{\mu \nu \rho}\,
\epsilon_{b_1b_2b_3} (R')_{b_1c_1}(R')_{b_2c_2} (R')_{b_3c_3} 
(\hat{n}_a)_{c_1} \partial_\nu (\hat{n}_a)_{c_2} 
\partial_\rho (\hat{n}_a)_{c_3}
$$
\begin{equation}
\,=\, (\det R') \, 
\oint_{\partial\theta} dS_\mu\,  \epsilon_{\mu \nu \rho}\, \hat{n}_a
\cdot (\partial_\nu \hat{n}_a \times \partial_\rho \hat{n}_a)\;,
\end{equation}
where we have used the property:
\begin{equation}
\epsilon_{b_1b_2b_3} (R')_{b_1c_1}(R')_{b_2c_2} (R')_{b_3c_3} \,=\,
(\det R') \, \epsilon_{c_1c_2c_3} \;.
\end{equation}
A similar relation holds if one changes the original canonical basis by a
constant rotation matrix.
What this proves is that it is possible to generate a monopole charge along
any color direction as long as one needs how to do that for, say, the third one. 

Thus, coming back to the discussion on the possible form of the kinetic
terms, they must be -at least in the symmetric phase-  compatible with the
local discrete gauge symmetry:
\begin{equation}
\lambda^a_{\mu} \rightarrow \lambda^a_{\mu} + \partial_\mu \omega^a,
\label{Abe-d}
\end{equation}
where $\omega^a$ is a {\em discontinuous\/} function taking values $\pm
\frac{g}{2}$ inside a three-volume $\vartheta^a$, and zero outside. 
Note also that in a phase only containing closed center vortices, a larger symmetry,
\begin{equation}
\lambda^a_{\mu} \rightarrow \lambda^a_{\mu} + \partial_\mu \varphi^a,
\label{Abe-c}
\end{equation}
for any smooth $\varphi^a$, is expected: in this case, the absence of monopoles would imply $\partial_\mu {\cal F}^a_{\mu}(C)=0$.
  
Then, the kinetic terms must have the global $SO(3)$ symmetry plus the local Abelian one. The simplest choice, which, in the effective
theory approach spirit we shall consistently adopt, is to minimally couple $\phi$
and $Q$ to $\lambda^a_\mu$ (note that these couplings do not have the local
$\vec{\lambda}(x) \rightarrow R(x) \vec{\lambda}(x) $ symmetry). 
Thus, the
structure of the kinetic term $K$ is as follows:
\begin{eqnarray}
K \;=\; K_\phi + K_Q \;,
\end{eqnarray}
where:
\begin{eqnarray}
K_\phi &=& \frac{1}{2} (\nabla_\mu \phi)^a (\nabla_\mu \phi)^a \nonumber\\
K_Q &=& \frac{1}{2} (\nabla_\mu Q)^{ab} (\nabla_\mu Q)^{ab}
\end{eqnarray} 
where $\nabla_\mu$ denotes the covariant derivative operator (consistent
with the symmetries mentioned above), which shall adopt a different
expression when acting on each one of the fields. Explicitly:
\begin{eqnarray}
 (\nabla_\mu \phi)^a  &=& \partial_\mu \phi^a - i g_\phi \, \lambda^b_\mu
\epsilon^{abc} \phi^c \nonumber\\
(\nabla_\mu Q)^{ab}  &=& \partial_\mu Q^{ab} - i g_Q \, \lambda^c_\mu
\epsilon^{acd} Q^{db}  + i g_Q \, Q^{ad} \epsilon^{dcb} \lambda^c_\mu \;,
\label{cova-d}
\end{eqnarray} 
where $g_\phi$ and $g_Q$ are constants. 

The global $SO(3)$ symmetry is evident, while by imposing $g_\phi=g_Q$, the effective action will display a non Abelian gauge symmetry, and the different phases 
for the ensemble of monopoles and center vortices will arise as different possible vacua when the system undergoes SSB.
Note also that as the field $\phi$ represents center vortices that in the case of Abelian configurations posses a 
magnetic charge $2\pi/g$, the natural choice is $g_\phi=2\pi/g$, which also matches the correct dimensions in eq. (\ref{cova-d}) as $[\lambda]=3/2, [g]=1/2$. 

Then, joining the different pieces and taking into account eq. (\ref{YM-1}), the following model, encoding a 
general ensemble of magnetic defects, can be proposed,
\begin{equation}
 {\cal L}_{{\rm eff}}= {\cal L}_{v,m}+\frac{1}{2} 
\lambda^a_\mu \lambda^a_{\mu}+i \lambda^a_{\mu} {\cal F}^a_{\mu}({\cal A}),
\label{eff-mod}
\end{equation}
\begin{equation}
 {\cal L}_{v,m}=K_Q+ K_{\phi}+V_\phi + V_Q  +V_I.
 \label{Lvm}
\end{equation}
We would like to underline that according to the discussion at the end of \S \ref{sec:nonab}, and beginning of \S \ref{sec:effective}, the symmetry 
displayed by the second and third terms in eq. (\ref{eff-mod}), namely a transformation $\vec{\cal A}^{\tilde U}$, accompanied by the local $SO(3)$ rotation of 
$\lambda^a_\mu$, is not the gauge symmetry that operates on $\vec{A}_\mu$. Therefore, the noninvariance of ${\cal L}_{v,m}$ under local $SO(3)$ rotations of
$\lambda^a_\mu$ is not an explicit breaking of the gauge symmetry in our effective model. Only in the trivial sector $\vec{\cal A}^{\tilde U}$
may be associated with a gauge transformation, in other words, our model refers to the interaction of effective fields, parametrizing 
a general ensemble, with effective gluons represented by $\vec{{\cal A}}_\mu$.

\section{Phase structure}\label{sec:phase}

In order to analyze the possible phases of the model, it is necessary to
study all the possible scenarios regarding both the $\phi$ and $Q$
dependent potentials, $V_\phi$ and $V_Q$, as well as the interaction $V_I$.
This will yield information about the possible translation invariant
configurations that will determine the properties of each phase. Non
translation invariant configurations, on the other hand, are important to
understand the mechanism driving the phase transitions between them. Of
course, that will require the inclusion of the derivative terms into the
game.

Thus we consider the minima of
\begin{equation}
V_{ T} \;=\; V_\phi + V_Q  + V_I\;.
\end{equation}
This analysis is greatly simplified if we note that $Q$ can always be
diagonalized by a similarity transformation $Q=R^T D R$, with 
\begin{equation}
D=\left( \begin{array}{ccc}
 -\frac{q}{2}-\frac{\eta}{2} & 0 &0  \\
0 &  -\frac{q}{2}+\frac{\eta}{2} &0 \\
 0 & 0 & q \\
\end{array}\right) .
\end{equation}
Defining $R\phi =\psi$, the potential $V_T$ adopts the form,
\begin{equation}
V_T \;=\; \frac{A}{2} \delta + \frac{B}{3} \Delta + \frac{C}{4} \delta^2 + \frac{D}{5} \delta \Delta + \frac{E}{6} \Delta^2
+\frac{\mu^2}{2} \psi^T \psi +\frac{\lambda}{4}(\psi^T \psi)^2 + \zeta \psi^T D\, \psi ,
\label{diag-V}
\end{equation}
\begin{eqnarray}
\delta &=& (3q^2+\eta^2)/2 \nonumber \\
\Delta &=& 3q(q^2-\eta^2)/4 \nonumber \\
\psi^T D\, \psi &=& -\frac{q}{2}(\psi_1^2+\psi_2^2) +\frac{\eta}{2}
(\psi_2^2 -\psi_1^2) + q \psi_3^2 \;.
\end{eqnarray}
Here, the term $\delta^3$ that was present in eq. (\ref{VQ})
has been discarded, as it does not modify the qualitative structure of the minima \cite{degennes-crys}.

Now, to find the minima of the potential, we will suppose that the chain of
spontaneous breaking of the symmetries is dominated by the monopole sector.
Concretely, this approximation amounts to finding the minima of $V_Q$, and
using the configurations $q_0, \eta_0$ that $Q$ adopts in those minima as a
fixed background where we look for the vortex field configuration that
minimizes the remaining potential. Then, the whole space of minima is generated by 
means of $R$-rotations of the former.

The minima of $V_Q$ are determined by:
\begin{equation}
\partial_q V_Q\big|_{q_0,\eta_0} = 0 \;\;,\;\;\;
\partial_\eta V_Q\big|_{q_0,\eta_0} = 0 \;\;,
\end{equation}
plus the usual conditions on the second derivatives. We will consider $CE>6D^2/25$, and will follow the discussion in \cite{degennes-crys},
where the different kinds of minima are obtained by varying $A$ and $B$. Changing the independent variables $q$ and $\eta$, the region
$\delta^3 \geq 6 \Delta^2$ is mapped, and the strict inequality occurs for
$\eta \neq 0$. Then, the points obtained by simply minimizing with respect
to $\delta$, $\Delta$ as independent variables (in this case the potential
contains a positive definite quadratic form) can only correspond to
$\eta_0\neq 0$. Otherwise, the potential must be minimized with the
constraint $\delta^3 =6 \Delta^2$, in which case two different situations
are obtained, $q_0=0, \eta_0=0$ or $q_0 \neq 0, \eta_0=0$ (when $CE< 6D^2/25$, only the two latter
possibilities can be realized).
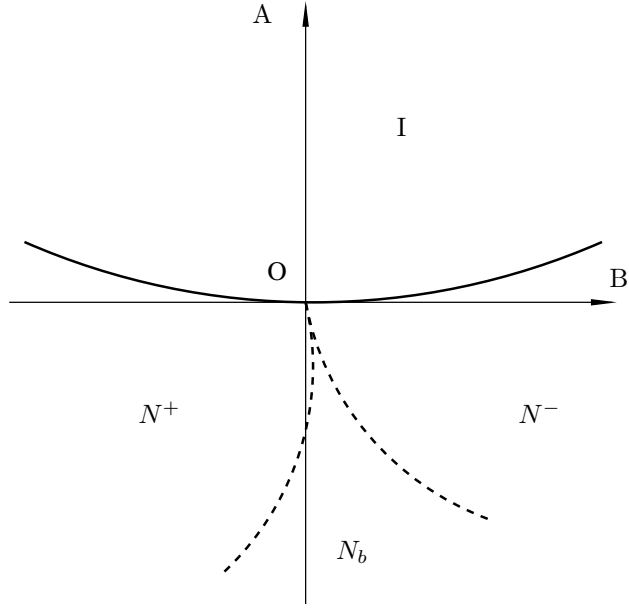
\begin{figure}
\begin{center}
\begin{picture}(0,0)%
\includegraphics{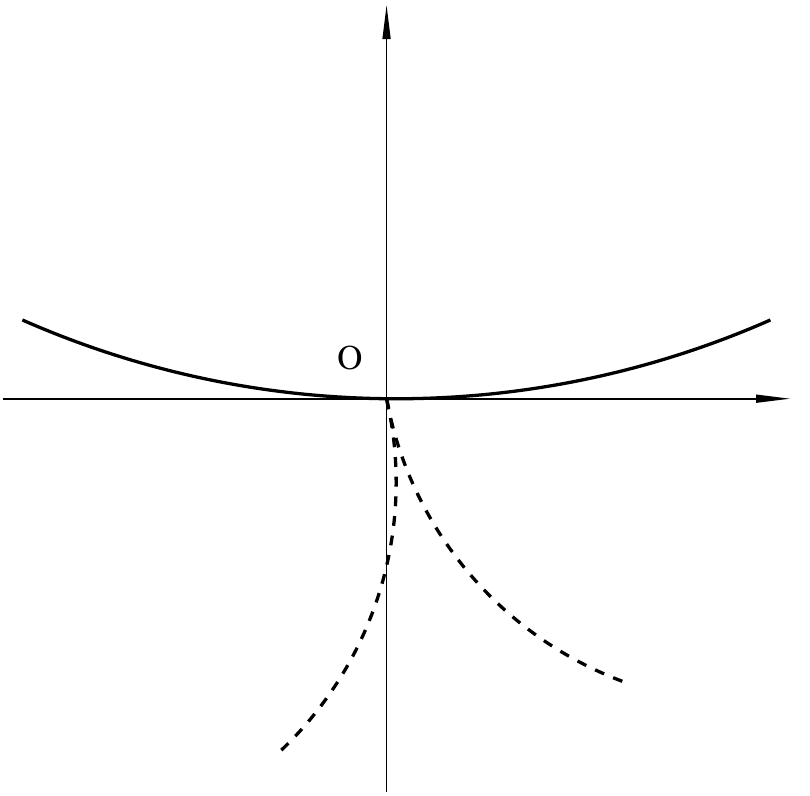}%
\end{picture}%
\setlength{\unitlength}{4144sp}%
\begingroup\makeatletter\ifx\SetFigFont\undefined%
\gdef\SetFigFont#1#2#3#4#5{%
  \reset@font\fontsize{#1}{#2pt}%
  \fontfamily{#3}\fontseries{#4}\fontshape{#5}%
  \selectfont}%
\fi\endgroup%
\begin{picture}(3689,3624)(484,-3223)
\put(1261,-2131){\makebox(0,0)[lb]{\smash{{\SetFigFont{10}{12.0}{\rmdefault}{\mddefault}{\updefault}{\color[rgb]{0,0,0}$N^+$}%
}}}}
\put(3511,-2131){\makebox(0,0)[lb]{\smash{{\SetFigFont{10}{12.0}{\rmdefault}{\mddefault}{\updefault}{\color[rgb]{0,0,0}$N^-$}%
}}}}
\put(2431,-2941){\makebox(0,0)[lb]{\smash{{\SetFigFont{10}{12.0}{\rmdefault}{\mddefault}{\updefault}{\color[rgb]{0,0,0}$N_b$}%
}}}}
\put(2791,-421){\makebox(0,0)[lb]{\smash{{\SetFigFont{10}{12.0}{\rmdefault}{\mddefault}{\updefault}{\color[rgb]{0,0,0}I}%
}}}}
\put(1936,254){\makebox(0,0)[lb]{\smash{{\SetFigFont{10}{12.0}{\rmdefault}{\mddefault}{\updefault}{\color[rgb]{0,0,0}A}%
}}}}
\put(4051,-1321){\makebox(0,0)[lb]{\smash{{\SetFigFont{10}{12.0}{\rmdefault}{\mddefault}{\updefault}{\color[rgb]{0,0,0}B}%
}}}}
\end{picture}%
\end{center}
\caption{A-B phase diagram for the monopole sector, when $CE>6D^2/25$, $D<0$.}
\label{diag} 
\end{figure}

Then, the $\psi$-field minima follow from the study of the `effective
potential' ${\cal V}(\psi)$, defined by:
\begin{equation}
{\cal V}(\psi) = V_\phi + V_I(q_0,\eta_0;\psi) \;, 
\end{equation}
and which explicit form is:
\begin{eqnarray}\label{eq:veff}
{\cal V}(\psi) &=& \frac{1}{2}\big[\mu^2 - \zeta (q_0 + \eta_0) \big] \psi_1^2 +
\frac{1}{2}\big[\mu^2 - \zeta (q_0 - \eta_0) \big] \psi_2^2 \nonumber\\
&+& \frac{1}{2}(\mu^2 + 2 \zeta q_0) \psi_3^2 +
\frac{\lambda}{4}(\psi_1^2 + \psi_2^2 + \psi_3^2)^2 \;.
\label{Vp}
\end{eqnarray}

It is now clear what kind of vacua may emerge, depending on the relative
values of the parameters. We first note that stability requires $\lambda
\geq 0$. For $D<0$, the monopole phase diagram is that of
fig.~\ref{diag} \cite{degennes-crys}. If the parameters $A$, $B$ are
initially in region I, we have $Q=0$ ($q_0=\eta_0=0$), and the effective
vortex potential results, ${\cal V}^{\rm I}(\psi) = \frac{\mu^2}{2}  \psi^2
+ \frac{\lambda}{4}(\psi^2)^2$. Then, if $\mu^2 \geq 0$, the minimization
gives $\psi=0$. With this vacuum, $S_{v,m}$ displays a non Abelian gauge
symmetry, much larger than the Abelian symmetries in eqs. (\ref{Abe-d}),
(\ref{Abe-c}), typically obtained when monopoles and center vortices are
present.  Therefore, this phase represents a situation where monopoles and
center vortices do not proliferate (deconfining phase). Still in the $Q=0$
phase, but with $\mu^2<0$,  the system undergoes SSB leaving an Abelian 
symmetry.  
If the mass scale generated for the off-diagonal fields is large, they will be suppressed and then the 
effective theory will essentially be invariant under Abelian gauge transformations of the form, 
\[
\vec{\lambda}_{\mu}\cdot \hat{\phi}_0 \rightarrow \vec{\lambda}_{\mu}\cdot \hat{\phi}_0 + \partial_\mu \varphi .
\]
Thus, recalling eq. (\ref{Abe-c}), this phase describes an ensemble of closed center vortices.

Now, in order to continue the analysis, it is convenient to define a
complex field $V=\frac{1}{\sqrt{2}}(\psi_1 +i \psi_2)$, and rewrite eq.
(\ref{Vp}) in the form (we consider $\zeta<0$), 
\begin{eqnarray}\label{eq:veff1} 
{\cal V}(\phi) &=&
(\mu^2 + |\zeta| q_0  )\, \bar{V}V +\frac{1}{2} |\zeta| \eta_0\,
(V^2+\bar{V}^2) \nonumber\\ &+& \frac{1}{2}(\mu^2 - 2 |\zeta| q_0)\,
\psi_3^2 + \lambda \left( \bar{V}V + \frac{1}{2} \psi_3^2\right)^2 \;.
\end{eqnarray}
When $A$ is lowered from positive to negative values, after a first order
transition, we will enter the uniaxial nematic phase $N^+$ ($q_0>0$) or the
$N^-$ ($q_0<0$), depending on whether $B<0$ or $B>0$. These phases are
characterized by $\eta_0=0$. In what follows, to simplify the analysis, we
will suppose $\mu^2 > 0$.  Then, if we enter the $N^-$ phase, after a discontinuous transition, the
effective potential ${\cal V}^-(\psi)$ will be minimized by $\psi_3=0$. In the monopole sector, 
the vacuum will be invariant under rotations around the third axis, while in the $V$-sector
this symmetry will undergo a U(1) SSB or not, depending on the sign of 
$(\mu^2 + |\zeta| q_0  )$. In addition, the $N^-$ phase will induce a mass of order $q_0^2$ for the charged dual vector fields 
$\lambda_\mu^1$ and $\lambda_\mu^2$, originated from the covariant derivative of $Q$ in eq. (\ref{cova-d}). If we assume this mass 
to be large when compared with the other mass scales in the problem, these dual vector fields will become suppressed.

If we further diminish $A$, after a second order phase transition, we will eventually reach the biaxial phase $N_b$ where $\eta_0
\neq 0$. As this transition is continuous, and we are approaching from the
$N^-$ phase, we will start with $\psi_3$ and $\eta_0$ small. In the $N_b$ phase, the $U(1)$ symmetry of the effective action 
in the former  $N^-$ phase will be broken to a discrete one under $\pi$-rotations along the third axis. 
Again, in the monopole sector the vacuum is invariant, while in the vortex sector it will display SSB of the discrete $\pi$-rotations depending
on the sign of $(\mu^2 + |\zeta| q_0)$. When this quantity is negative, at the minima, the $V$ field can take a pair of values $V_0$,
$-V_0$ connected by a $Z(2)$ symmetry.  That is, the obtained effective
potential coincides with the confining phase of the vortex model introduced
by t' Hooft, relying on the possible nontrivial vortex correlators in the
initial theory. In this phase, the spontaneous $Z(2)$ symmetry breaking
leads to domain walls attached to Wilson loops, thus providing an area law.
Still in the $(\mu^2 + |\zeta| q_0)<0$ case, in the intermediate $N^-$ phase, the vacuum no longer displays the Abelian
symmetry present in the initial phase, where center vortices are only
closed objects, nor the discrete symmetry of the last phase, typical of
open center vortices whose endpoints are joined in pairs to monopole-like objects that proliferate. From this
perspective, we speculate that the $N^-$ phase might be associated with one
where monopoles and antimonopoles are still bound in pairs. 

\section{Conclusions}\label{sec:conclu}

We have constructed a novel non Abelian effective model for $SU(2)$ QCD in
Euclidean three-dimensional spacetime that allows for the description of a
phase diagram with a rich structure.  The construction is based on a
special parametrization of the gauge field configurations $\vec{A}_\mu$ in
terms of a vector field $\vec{{\cal A}}_\mu$, representing a topologically
trivial sector of smooth fluctuations, and a local color frame $\hat{n}_a$
containing defects, the nontrivial sector describing monopoles and thin
center vortices.  The frame can be written as a local $SO(3)$ rotation $R$
of the canonical basis $\hat{e}_a$, which can be also expressed in the form
$R=R(S)$, where $S$ is in the fundamental representation. 

This parametrization is used to write the Yang-Mills action, what defines
the weight assigned to each configuration.  On the other hand, as in any
non-perturbative definition of the functional integration measure in a non
Abelian gauge theory, one is faced with the usual stumbling blocks, related
to the Gribov problem. We do not attempt to tackle this problem; rather,
since we use the functional integral just as a guide for the subsequent
derivation of the effective model, we use instead a definition of the
measure which: (a) reduces to the proper one for topologically trivial
configurations and (b) is consistent with (although not uniquely determined
by) the properties of the gauge field parametrization used. 

The next step in the construction of the effective model proceeds with the
introduction of an auxiliary field $\vec{\lambda}_\mu$ that linearizes the
Yang-Mills action, and the incorporation of a phenomenological weight $S_d$
that senses the geometry of the defects. It is at this point where the real
reduction to an effective theory is implemented. Indeed, the symmetries are
identified here, for a given classification of defects, what allows us to
construct an effective model. 

If $S_d$ were nullified (thin objects), the partition function for the
sector of defects should be invariant under local $SO(3)$ rotations of
$\vec{\lambda}$, as they could be absorved by a frame redefinition,
transforming $S$ under right multiplication by an appropriate regular
$SU(2)$ matrix $\tilde{U}^{-1}$. In the Yang-Mills partition function, the
symmetry should also be accompanied by the transformation $\vec{\cal
A}^{\tilde{U}}_\mu$. However, this symmetry is the one associated with the
many different ways a given gauge field $\vec{A}_\mu$ containing thin
defects can be decomposed, so that it is expected to be broken as soon as
center vortices become thick. Alternatively, this could be seen as the
noninvariance of the effective phenomenological action $S_d$ under local
frame rotations, only leaving a global $SO(3)$. 

An interesting point is that in order to guide the construction of the
effective model for the ensemble integration, not only the global $SO(3)$
symmetry is important but also a new symmetry comes into play. At least in
the symmetric phase, due to the topological structure of monopoles, the
model should be invariant under a local discrete gauge symmetry.  This led
us to propose a non Abelian model describing the interaction of the natural
order parameters for monopoles and center vortices with the
effective gluon field $\vec{\cal A}_\mu$. As center vortices can be
attached in pairs to the non Abelian monopoles, the corresponding order
parameters are  given by fields $\phi$ and $Q$, carrying isospin one and
two, respectively. The effective character of the gluons is due to the fact
that gauge transformations of the Yang-Mills fields $\vec{A}_\mu$ act as a
left multiplication of the $S$ sector, leaving $\vec{{\cal A}}_\mu$
invariant. 

 The effective model we introduced exhibits a rich phase diagram.  For instance,
the monopole sector of the effective potential depends on two invariants,
$\delta = Tr\, Q^2 $, $\Delta= Tr\, Q^3$.  If this sector is supposed to
dominate the transitions, the phase diagram inherits, by construction, some
of the properties found in liquid crystals. In this case, if the quadratic
form in the quantities $\delta$ and $\Delta$ is positive definite, and the
coefficient of the linear $\Delta$-term is positive, an interesting chain
of phase transitions is obtained. 

Initially, when the coefficient of the linear $\delta$-term ($A$) is varied
from positive to negative values, a first order transition from the
isotropic deconfining phase to a uniaxial monopole condensate takes place.
In this process, in the vortex sector, the ``third'' component becomes
suppressed, while the other two components can be
arranged as an Abelian complex vortex field $V$ displaying $U(1)$ SSB. In this example, the vortex mass scales
have been supposed to be negligible when compared with those generated in
the monopole sector.  The further reduction of $A$ produces a second order
phase transition, and the monopole condensate becomes biaxial. Here, center
vortices are left in a global $Z(2)$ SSB phase, thus making contact with
the 't Hooft vortex model, and arriving to the confining phase expected in
$3D$ Yang-Mills theories.

\section*{Acknowledgements}
The Conselho Nacional de Desenvolvimento Cient\'{\i}fico e Tecnol\'{o}gico
(CNPq-Brazil) and PROPPi-UFF are acknowledged for the financial support.

\noindent
\mbox{C.\ D.\ F.\ } ac\-knowl\-edges financial support from CONICET and UNCuyo. 


\end{document}